# Controlling the dynamics of a bidimensional gel above and below its percolation transition

D. Orsi[1], B. Ruta[2], Y. Chushkin[2], A. Pucci[3], G. Ruggeri[3], G. Baldi[4], T. Rimoldi[1] and L. Cristofolini[1*]

1 Department of Physics and Earth Sciences, University of Parma, Viale Usberti 7/A, 43124 Parma, Italy
2 European Synchrotron Radiation Facility, B.P. 220, F-38043 Grenoble, France
3 Department of Chemistry and Industrial Chemistry, University of Pisa, Via Risorgimento 35, 56126 Pisa, Italy
 4 IMEM-CNR Institute, Parma Science Park, I-43124 Parma, Italy

*luigi.cristofolini@unipr.it

The morphology and the microscopic internal dynamics of a bidimensional gel formed by spontaneous aggregation of gold nanoparticles confined at the water surface are investigated by a suite of techniques, including grazing-incidence x-ray photon correlation spectroscopy (GI-XPCS). The range of concentrations studied spans across the percolation transition for the formation of the gel. The dynamical features observed by GI-XPCS are interpreted in view of the results of microscopical imaging; an intrinsic link between the mechanical modulus and internal dynamics is demonstrated for all the concentrations. Our work presents, to the best of our knowledge, the first example of a transition from stretched to compressed correlation function actively controlled by quasistatically varying the relevant thermodynamic variable. Moreover, by applying a model proposed time ago by Duri and Cipelletti [A. Duri and L. Cipelletti, Europhys. Lett. 76, 972 (2006)] we are able to build a novel master curve for the shape parameter, whose scaling factor allows us to quantify a "long time displacement length". This characteristic length is shown to converge, as the concentration is increased, to the "short time localization length" determined by pseudo Debye-Waller analysis of the initial contrast. Finally, the intrinsic dynamics of the system are then compared with that induced by means of a delicate mechanical perturbation applied to the interface.

61.05.cf – X-ray scattering (including small-angle scattering)
83.80.Kn       Physical gels and microgels
68.18.Fg       LB: Liquid thin film structure: measurements and simulations

## *1. Introduction*
Films formed by nanoparticles confined at an interface, either separating liquid from air or two immiscible liquid phases, are extensively studied [1] due to their importance in a wide range of systems, from the stabilization of emulsions and of foams [2] to the formation of photonic or plasmonic membranes [3,4]. Films made by noble metal nanoparticles in particular have recently been shown to form mechanically robust, free-standing membranes [5,6] which can be important for the fabrication of high-efficiency solar cells and displays.

Unlike the case of 3D colloidal suspensions, which often can be described reasonably well using Derjaguin–Landau–Verwey–Overbeek theory (DLVO), the description of particles at interfaces requires the consideration of interfacial deformations as well as interfacial thermal fluctuations, as complex intercolloidal forces act within these 2D assemblies, which also depend on the contact angle of the individual colloids, with possible long ranging effects. Moreover, hydrodynamic coupling between the 2D films and the bulk of the surrounding fluids complicates the measurement and the modeling of surface dynamic properties. For all these reasons, the theoretical comprehension of interfacial colloidal systems still represents an outstanding problem in the modern theory of colloidal interactions [7].

Langmuir monolayers of colloids at the air-water interface constitute a model system commonly used to investigate the structure and properties of such interfacial systems, recently applied also to the study of films of gold nanoparticles [8].

We remark that accurate structural and mechanical characterizations, with a quasi-statically increase of the concentration, are peculiar of the 2D geometry; a comparable analysis on 3D systems would be much more difficult. Interfacial stability may become an issue when the concentration is increased. Structural investigations by grazing incidence X-ray scattering have been employed to elucidate the structure of deposited Langmuir Blodgett [9] and unsupported Langmuir monolayers [10]. In favorable cases, the contact angle of individual particles could be estimated by accurate modeling of the x-ray reflectivity curve [11,12].

In these films, high concentration regimes may present temporal and spatial dynamical heterogeneities, in conjunction with the arising of slow dynamics often leading to dynamical arrest and to ageing phenomena. Such heterogeneities reflect the complexity of the structural arrangements and interactions among the colloids. As a matter of fact, dynamical heterogeneities of different kind are the fingerprints of the different models that over the years have been put forward to describe dynamical arrest. Their detailed quantification is therefore a crucial step in understanding the dynamics of systems approaching the arrested state [13].

This task can be accomplished by measuring high order correlation functions, e.g. to characterize the non-Gaussian dynamics of arrested systems [14,15]. In a previous work, we have been able to identify heterogeneities by grazing incidence X-ray photon correlation spectroscopy (GI-XPCS) [16] in a Langmuir polymeric monolayer [17] and in the highly compressed phase of a 2D gel generated as a Langmuir monolayer of gold nanoparticles [18]. In the former case, heterogeneous dynamics could be observed clearly in the photo-induced fluid phase of the polymer, while in the latter case, heterogeneity was found to be growing with increasing surface compression, and the quality of the data was good enough to allow the first experimental determinations of the fourth-order temporal correlation function in an arrested system by GI-XPCS. The morphology of the same system has been investigated in detail by microscopy techniques, revealing a complex structure with features on a hierarchy of different sizes following a Levy distribution [19]. The evolution of the statistical properties of the structured network, as a function of the externally controlled density, can be related to the corresponding evolution of the mechanical properties. Three steps can be identified in the gel formation: a first step occurs in the incubation time, in which the individual nm-sized gold nanoparticles aggregate to form quasi 1D structures of typical length of a few microns (flocculation), in a fashion similar to that reported in [20]. The second step occurs in the first stages of the compression, with the growth of the branched structure, finally yielding to the onset of the infinite percolative cluster, which is related to the building of the mechanical elastic modulus. Percolating transition occurs around $\Phi = 31.9\%$ [21].

It is well known that fluctuation dynamics are strictly related to mechanical moduli, via generalized fluctuation-dissipation relations [22]. This has been extensively exploited for the characterization of the rheology of interfacial layers: the measurement of their mechanical properties is a delicate task that can be performed by a number of techniques, which can be broadly divided into active and passive ones. In the former case, a deformation is applied from outside, as in the oscillating barriers technique [23] and in the interfacial shear rheology, in which a magnetized needle is used to induce shear deformation [24,25]. In the latter case, mechanical properties are deduced from the spontaneous fluctuations recorded e.g. by dynamic light scattering or by XPCS experiments [17,26,27]. A priori, the combination of the results of such measurements to yield a comprehensive picture is complicated by the different space scales involved, spanning from the centimeter for the ISR down to fractions of a micron in the case of XPCS. However, in most cases, those macroscopic and microscopic quantities have been found to agree, as e.g. in floating polymeric layers [17,26]. A counterexample of this is found e.g. even in simple phospholipid monolayers; their macroscopically liquid and mainly viscous phase may exhibit a complex and mainly elastic response when probed on the micron-scale [28].

In the present paper we report a detailed study of the spontaneous fluctuation dynamics of a Langmuir monolayer of gold nanoparticles floating on the water surface, together with the relation to its structure and to its mechanical response. This is done, exploiting a peculiaritiy of the Langmuir technique, varying the surface concentration from well below to well above the threshold for the formation of a percolative cluster. The sample, its preparation and the experimental methods used are described in details in section 2. We start with a characterization of the system by different microscopy techniques such as in situ epi-microscopy of the Langmuir monolayer and SEM imaging of the monolayers transferred onto solid substrate; next, we report on the measurement of the mechanical moduli of the film (section 3.1). Then, by means of GI-XPCS we focus on the spontaneous fluctuation dynamics and the connections to mechanical moduli; we investigate the evolution while crossing the percolation threshold (section 3.2). We discuss the changes in the dynamics and their connection with structure and mechanical moduli in light of different theories. Finally, we report on the effects that an external mechanical perturbation exerts on the fluctuation dynamics of the system, with particular regard to transient states (section 3.3).

## *2. Sample and Methods*

### 2.1 Sample

Nanoparticles have been synthesized as follows, by the two-phase *Brust* method [29,30]. In brief, $0.400\ g$ ($1.02\ mmol$) of hydrogen tetrachloro-aurate(III) trihydrate have been dissolved in $30\ mL$ of deionized water. The solution has been then shaken in a separatory funnel with $80\ mL$ of toluene solution containing $2.00\ g$ ($3.56\ mmol$) of tetra-n-octylammonium bromide (TOAB). The toluene phase has been then recovered and combined with $0.018\ mL$ ($0.015\ g$, $0.07\ mmol$) of dodecyl mercaptane. A freshly prepared aqueous solution of sodium borohydride ($25\ mL$, $0.386\ g$) has been slowly added under vigorous stirring. After further stirring for 3 hours, the organic phase has been separated, concentrated to $10\ mL$ and mixed with $70\ mL$ of ethanol. The mixture has been cooled overnight at $-20\ °C$ and the dark precipitate then recovered by filtration. The crude product has been then purified by soxhlet extraction with acetone as cleansing solvent to remove all the unbound free thiol and residual TOAB impurities, thus making GNP fully soluble in organic solvents. The average diameter of GNP has been estimated to be $8\ nm$ by DLS analysis with a $6\ nm$ diameter gold nucleus (TEM analysis) [31].

The Langmuir monolayer has been prepared as follows: nanoparticles have been dispersed in chloroform, $50\ mg$ in $10\ mL$, thus obtaining a $5\ mg/mL$ suspension used for long term storage in a refrigerator.

Before each experiment, a small batch of less concentrated suspension have been prepared by dilution of the stock solution in hexane, to reach the final concentration of $0.2\ mg/mL$. This suspension has been slowly spread at the air/water interface using a $100\ \mu L$ Hamilton syringe; the tip of the syringe's needle was kept in contact with the water surface during the whole process. A total volume of $2.5\ mL$ has been spread on the Langmuir trough used in GI-XPCS experiments: a spreading time of 2 minutes was used for the spreading of each syringe. A 20 minute waiting time was scheduled to allow solvent evaporation and to achieve the equilibration of the sample.

During the experiments the water subphase was kept at constant temperature ($18\ °C$), by means of water circulation from a *Lauda* thermal bath through the Langmuir trough's basement.

## 2.2 Rheology measurements

The mechanical response of the 2D system has been completely characterized, by measuring both the compression ($\varepsilon$) and shear ($G$) complex moduli as a function of surface concentration. This was accomplished by using two complementary techniques operating -on the macroscopic scale- on the same temporal scale: the *oscillating barriers* and the *oscillating needle* techniques.

The oscillating barriers technique exploit the difference in the surface pressure measured by perpendicular Wilhelmy plates when a dense Langmuir film is subject to uniaxial compression [23,32]. The experimental setup consisted of a Langmuir trough (custom made, size $10 \times 40\ cm^2$, with KSV Nima mechanics) equipped with two identical Wilhelmy balances, which monitor the evolution of the surface tension (and hence the phase and amplitude of their oscillations $\Delta\Pi_{||}$ and $\Delta\Pi_\perp$) while the area available to the film is oscillated by the synchronized movement of the two barriers, operated at a typical frequency of $0.5\ Hz$.

From the area $A_0$, its oscillation $\Delta A$ and the phase lag $\vartheta$ we obtain the complex mechanical moduli:

$$G = G' + iG'' = A_0 \left(\frac{\Delta\Pi_{||}}{\Delta A} - \frac{\Delta\Pi_\perp}{\Delta A}\right) e^{i\vartheta} \text{ eq.(1)}$$

$$\varepsilon = \varepsilon' + i\varepsilon'' = A_0 \left(\frac{\Delta\Pi_{||}}{\Delta A} + \frac{\Delta\Pi_\perp}{\Delta A}\right) e^{i\vartheta} \text{ eq.(2)}$$

This analysis assumes linearity in the response of the system to the external perturbation. This assumption was observed to hold in the diluted regimes, while at concentrations significantly higher than percolation, we observed important nonlinearities in the compression response even for the smallest amplitude of perturbation that was applicable. For this reason the oscillating needle technique, which imposes a pure shear perturbation, was employed to characterize the latter regime. The oscillating needle rheometer employed is an adapted version of the instrument proposed years ago by Fuller and coworkers [24] and is described in detail in [25]. In brief, a stainless steel needle (1cm long, 0.3mm radius) magnetized to saturation is incorporated in the Langmuir film and oscillates under the action of a suitable magnetic field gradient whose equilibrium position is oscillated in a sinusoidal manner. A fast CCD camera measures the amplitude and the phase of the oscillations of the needle. The mechanical shear modulus G is then obtained as follows:

$$G = G' + iG'' = \frac{\sigma}{\gamma} e^{i\vartheta} \text{ eq.(3)}$$

where $\sigma$ is the stress exerted by the needle, $\gamma$ is the resulting strain and $\vartheta$ is the phase difference between the oscillations of the magnetic field and of the needle.

### 2.3 GI-XPCS measurements

Multispeckle GI-XPCS experiments were performed at the beamline ID10 of the European Synchrotron Radiation Facility (ESRF) in Grenoble, France. An incident x-ray beam with an energy of $8\ keV$ ($\lambda = 0.155\ nm$) was selected from the undulator radiation by a Si(111) pseudo-channel cut monochromator scattering in vertical geometry (energy bandwidth $\Delta E/E \sim 10^{-4}$). The beam was then focused with Be compound lenses while higher harmonics were suppressed by two Si mirrors. The spatially coherent part of the incoming radiation was selected by using roller-blade slits opened to $10 \times 10\ \mu m$, placed $\sim 0.18\ m$ upstream of the sample. The parasitic scattering produced by the beam-defining slits was removed by carefully adjusting a set of guard slits a few cm upstream of the sample. The resulting incident flux on the sample was $10^9\ photons/s/100\ mA$. The beam was reflected to impinge on the liquid surface at a grazing incident angle of 0.137°, which is about 90% of the critical angle for total reflection on the water surface at this photon energy.

A custom-made Langmuir trough with a single moving barrier (maximum surface $418 \times 170\ mm^2$) was installed on the sample diffractometer. The trough was mounted on an active antivibration support and provided with a plastic enclosure, under which a helium atmosphere was created in order to minimize parasitic scattering from air and contemporarily reducing risks of beam-induced damage. KSV-Nima Wilhelmy balance and electronics were used to control the trough and measure surface pressure and trough's area during GI-XPCS measurements. The mechanical perturbation of the film was provided by a $4 cm$ long magnetized needle that was put in oscillation by two magnets inserted for this purpose. The magnets where suspended parallel to the water surface by means of a plastic support, with their dipoles mutually perpendicular. One of them was permanent, needed to define the equilibrium position and orientation of the needle, while the second, an externally controlled electromagnet, was used to rotate the needle thus providing the required perturbation to the film.

Two-dimensional x-ray scattering speckle patterns were recorder by using a photon counting area detector (Medipix, $256 \times 256$ pixels, $55\ \mu m$ pixel size, [33]) placed at a distance of $3.3\ m$ from the sample. The exposure time of the Medipix detector has been chosen to be long enough to warrant a reasonable S/N ratio, $10 - 100\ msec$ in our case. To avoid unnecessary irradiation of the sample, a fast shutter, synchronized with the area detector, was placed upstream of the sample. Sets of up to $20000$ images were collected for different sample concentrations with different exposure times. The intensity time correlation functions were calculated from the series of two dimensional images by software.

## *3. RESULTS AND DISCUSSION*

### 3.1 Morphology and mechanical response

Exploiting the peculiarity of the 2D geometry of a Langmuir experiment, we performed experiments covering a wide range of surface concentration $\Phi$. In our system, as $\Phi$ is increased, a 2D network of gold nanoparticles is formed. The process of formation of the network has several similarities with a percolation transition, as it has been thoroughly characterized in a microscopy experiment [21]. A sample of the resulting network, imaged by an inverted microscope, acquired with a 50X objective, is shown in figure 1a.

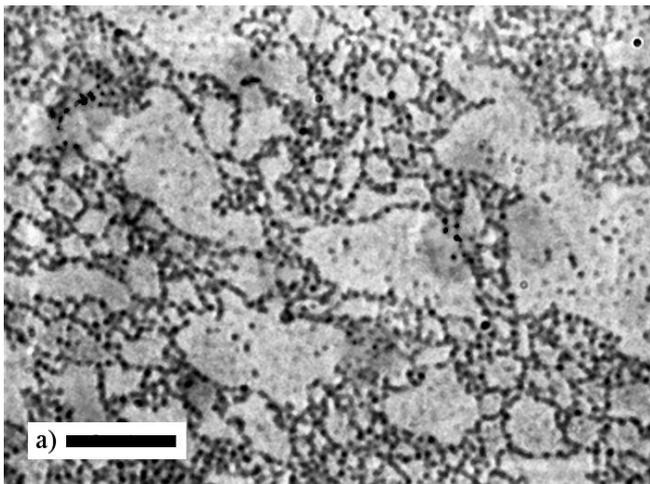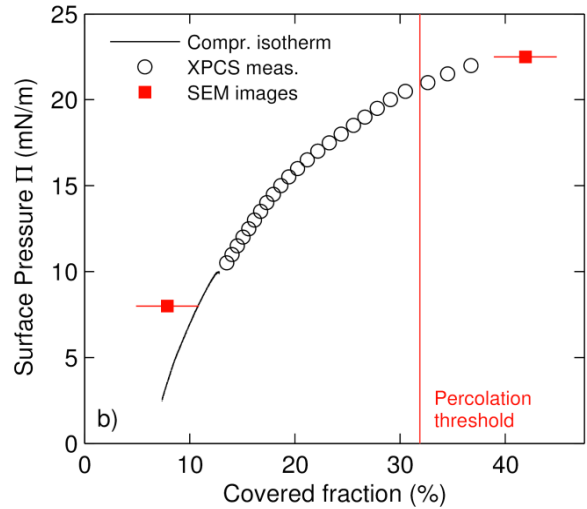

***Figure 1: At each surface pressure $\Pi$, the concentration of gold nanoparticles at the surface is monitored by imaging. a)*** *image of the GNP network at $\Phi = 29\%$, taken using an inverted microscope with a 50X objective (the black scale-bar is 200μm long)* ***b)*** *Surface pressure - concentration isotherm measured during the GI-XPCS experiment. GI-XPCS measurements were performed at the concentrations indicated by the empty circles. As a double-check, $\Phi$ was also directly measured by SEM imaging on two samples transferred on solid silicon substrate (red squares). Color online.*

Figure 1b shows the surface pressure – concentration $\Pi - \Phi$ isotherm recorded during the GI-XPCS experiment. The determination of the surface concentration $\Phi$ is very important for the following discussion, therefore it has been cross-checked by employing different techniques: In our laboratory, besides calculating $\Phi$ from the spread amount, as usual in Langmuir experiments, we also performed in-situ epi-microscopy and Brewster Angle Microscopy on the film floating at the air/water interface, in addition to Scanning Electron Microscopy (SEM) imaging of samples of film transferred onto Silicon substrate. The estimates of $\Phi$ arising from these investigation are all consistent with that calculated from the spread amount of nanoparticles [19,21].

For the GI-XPCS experiments reported here, the estimate from the spread amount has been cross checked ex-post by SEM imaging of samples of film transferred onto silicon substrate at selected values of $\Pi$. The results- reported in figure1b – show good agreement between the two determinations.

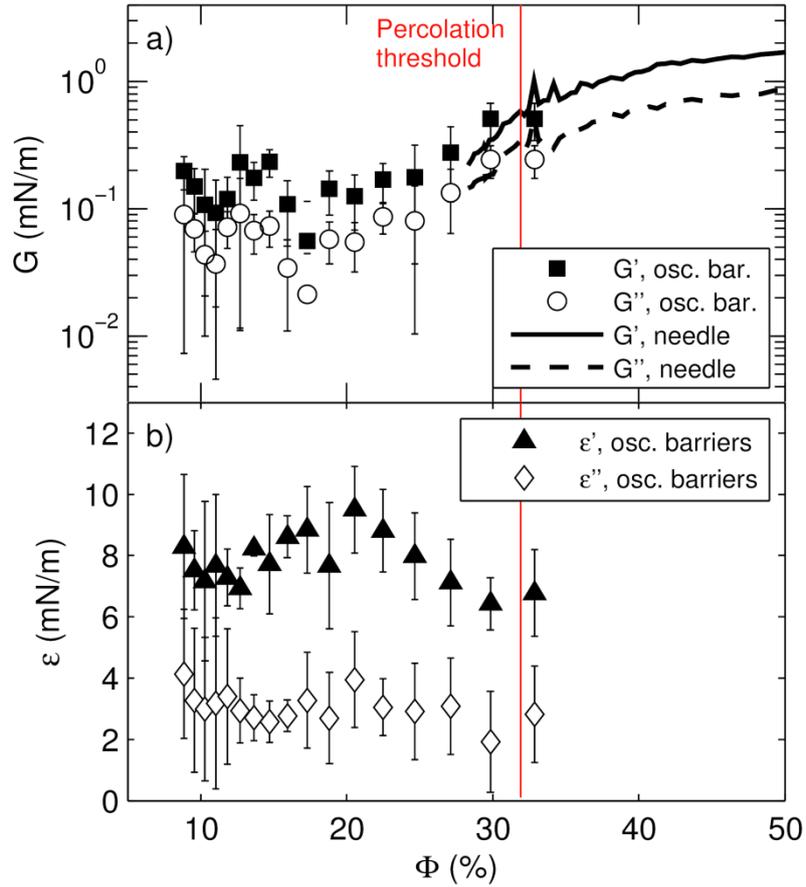

***Figure 2: Φ dependence of the mechanical properties of the GNP network during its formation process. a)*** *the network shows mainly elastic shear modulus, which increases with* Φ. ***b)*** *the compressibility stays constant up to the percolation threshold (vertical line).*

The evolution of the mechanical response above and below the percolation transition has been measured using the techniques discussed in the experimental section; results are reported in figure 2.

While it is reasonable to postulate that, at very dilute concentrations, a viscous/dissipative regime shall exist (which however may corresponds to extremely low values of the moduli, well below the sensitivity limits of the techniques used), in the range relevant to our case, both shear ($G$) and compression ($\varepsilon$) moduli are mainly real, indicating an elastic response, in agreement with previous results on the same system [19]. There is an important difference however between shear and compression responses. As shown in figure 2a, $G$ presents a transition between two different regimes: for $\Phi < 25\%$ it has a very small value (below 0.2mN/m, close to the sensitivity limit of the technique) while as the percolation threshold is approached, $G$ starts to increase, reaching a saturation value of the order of $1-2\ mN/m$, above which it enters a nonlinear regime, indicative of the intrinsic brittleness of this structure. On the contrary, as shown in figure 2b, the compression modulus $\varepsilon$ stays constant at all the concentrations below the threshold, and nonlinear effects dominate above this concentration.

### 3.2 Intrinsic dynamics probed by GI-XPCS

The dynamics of the GNP network have been measured by recording the X-ray intensity scattered from the sample in the grazing incidence diffraction experiment described in the experimental section, at several values of $\Phi$, in correspondence of different values of the components of the scattering vector in the directions parallel $(q_{\parallel})$ and perpendicular $(q_{\perp})$ to the air/water interface. The pixels of the 2D detector have been divided into square groups, each of which is labeled by its scattering vector components $(q_{\parallel}, q_{\perp})$. From the scattered intensity measured by each group of pixels as a function of time we calculate the intensity autocorrelation function

$$g^{(2)}(q_{\parallel}, q_{\perp}; t) = \frac{\langle I(t_0)I(t_0+t)\rangle_{t_0}}{\langle I(t_0)\rangle_{t_0}^2} \quad \text{eq.(4)}$$

In soft-matter experiments, the correlation functions are commonly described using the empirical Kohlrausch-William-Watts (KWW) exponential

$$g^{(2)}(q_{\parallel}, q_{\perp}; t) = A + \beta e^{-2\left(\frac{t}{\tau}\right)^{\gamma}} \quad \text{eq.(5)}$$

where $\tau$ is the relaxation time of the dynamics, $\beta$ is the contrast and $\gamma$, called shape parameter, is connected to the kind of dynamics that characterize the system, together with the dependence of $\tau$ from q. We briefly recall here that, in case of Brownian dynamics, $\tau \propto q^{-2}$ and $\gamma = 1$; pure ballistic motion shows $\tau \propto q^{-1}$ and $\gamma = 2$ [34], while more complex systems display some intermediate behavior, e.g. many arrested systems show a compressed shape often with $\tau \propto q^{-1}$.

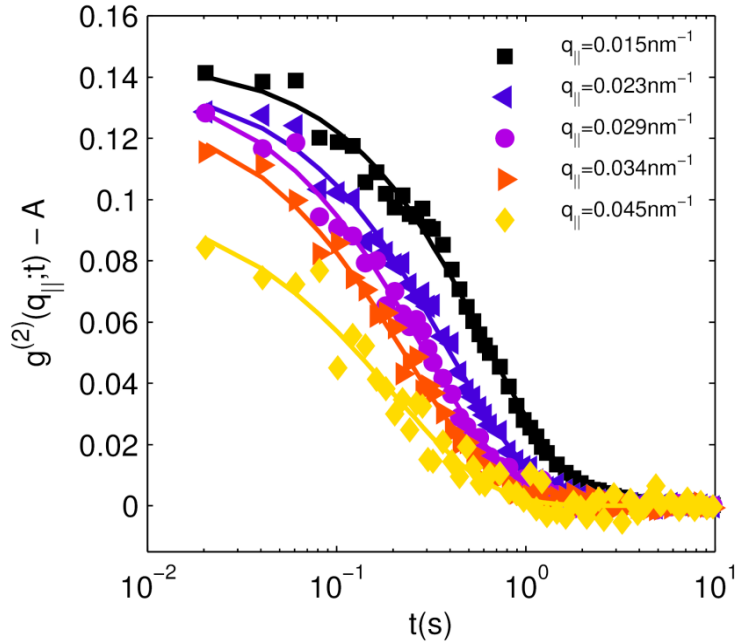

**Figure 3: correlation functions decay as KWW exponentials.** *Here, correlation functions measured at $\Phi = 15\%$ at several values of q are shown. Continuous lines are fits to KWW phenomenological law. Color online.*

Figure 3 reports data and fit models for correlation functions measured at $\Phi = 15\%$. The initial contrast of the correlation functions shows a clear dependence on q. When additional relaxation processes faster than the exposure time are present, they cause a sort of "blurring" of the speckle pattern. This implies a reduction of the contrast $\beta$ with respect to its theoretical value $\beta_0$; this reduction depends on q following a pseudo Debye-Waller decay [27] ruled by the localization length $r_{loc} = \sqrt{\langle u^2 \rangle}$:

$$\beta = \beta_0 \, exp\left(-\frac{q_{\parallel}^2 \langle u^2 \rangle}{3}\right) \quad \text{eq.(6)}$$

This is clearly the case for our data, as inspection of figure 3 shows that the initial contrast is well below the theoretical value which in our experimental geometry is approximately $\beta_0 = 0.2$. In figure 4a we report the q-dependence of $\beta$ together with the fits to equation 6. By this analysis, the localization length $r_{loc}$ can be extracted, as reported in figure 4b. The value of $r_{loc}$ decreases linearly with $\Phi$ at low concentration, reaching a value compatible with the particle's diameter (8 nm) at about 22%; while it stays constant at higher concentrations. The linear decrease is hallmarked by the black line in figure 4b.

Interestingly, 22% is close to the concentration range in which we found GNP flocculation into spherical aggregates that grow in size as concentration is increased [21], at higher concentration a second step occurs in the process of network formation: string-like aggregates appear on approaching the percolation transition.

Therefore, the decrease of $r_{loc}$ to 8nm upon increasing concentration, may be interpreted as if the fast single-nanoparticle motion becomes more and more hindered by the spatial constrains put by neighbors.

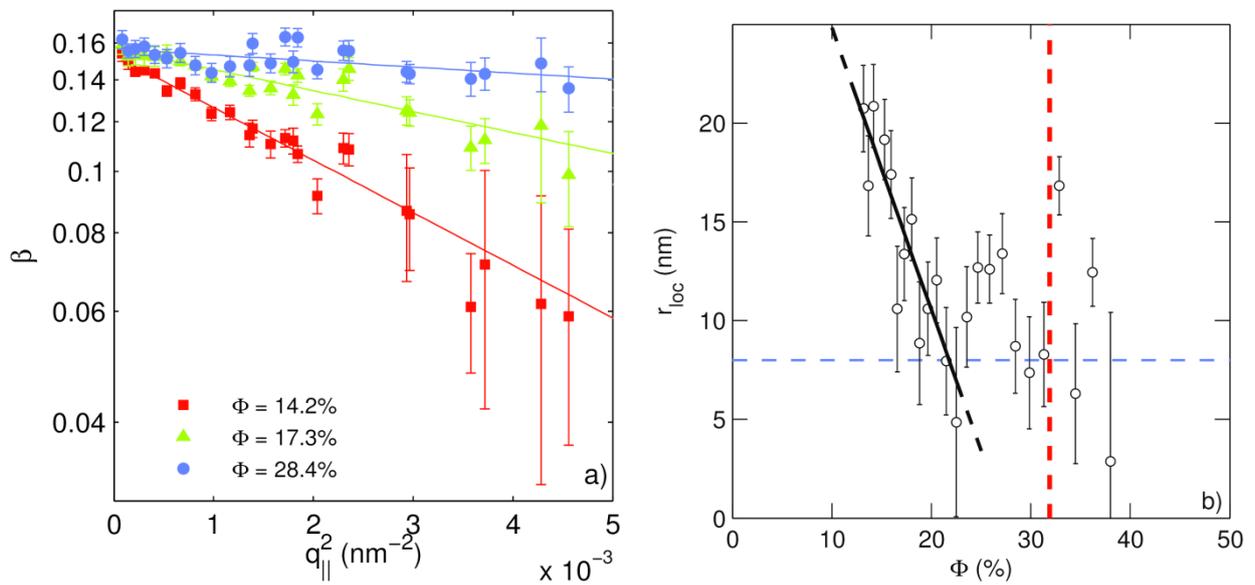

*Figure 4: the contrast decays with $q_{||}$ following a Debye-Waller dependence. a) Data for three different concentrations are plotted on a semi-logarithmic scale as a function of the squared scattering vector. Lines are fit to equation 6. b) localization length $r_{loc}$ as a function of concentration: increasing $\Phi$ it decreases to reach a value comparable with the size of an individual nanoparticle (8nm, dashed blue line). The vertical red line represents the percolation threshold. Color online.*

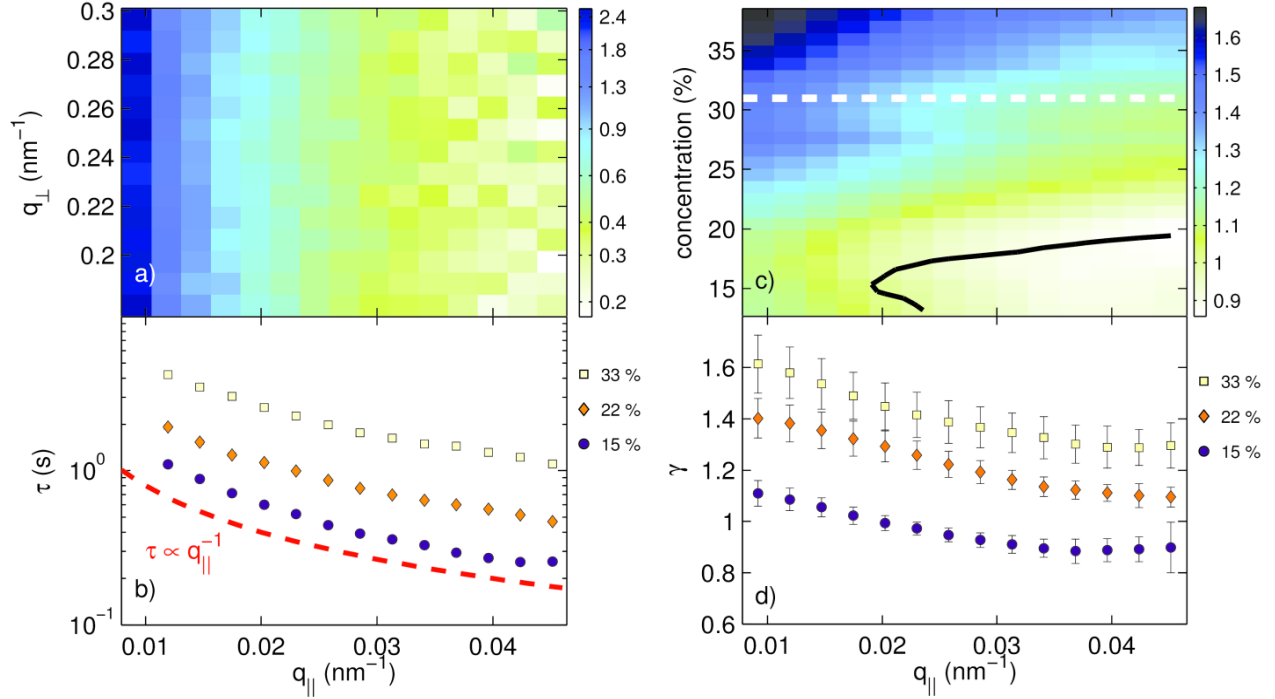

**Figure 5: both the relaxation time $\tau$ and shape exponent $\gamma$ decrease at large $q_{\|}$.**
*a) map of relaxation time $\tau$ as a function of the 2 components of the scattering vector, $q_{\|}$ and $q_{\perp}$, for $\Phi = 15\%$. Values of the color scale are in seconds. b) detailed dependence of $\tau$ from $q_{\|}$ measured at different concentrations ($\Phi = 15, 24, 34\%$). The $\tau \propto q_{\|}^{-1}$ behavior is also shown for comparison (dashed line) c) shape parameter $\gamma$ as a function of $q_{\|}$ and of $\Phi$. The black line represents the contour of $\gamma = 1$, i.e. of simple exponential decay, while the white dashed horizontal line indicates the critical concentration (see text) d) detailed $q_{\|}$ dependence of $\gamma$ for the same concentrations of panel b ($\Phi = 15, 24, 34\%$). Color online.*

In panel a) of figure 5 we report the values of $\tau$ obtained from the fits to the KWW relation of the correlation functions measured at $\Phi = 15\%$. In agreement with our previous investigations [18] we find that $\tau$ depends only on the parallel component of the scattering vector $q_{\|}$, which varies in the horizontal axis in the figure, and not on the perpendicular one, thus confirming that the dynamics is confined at the air water interface. Therefore, in all the subsequent analysis, we averaged the results along $q_{\perp}$ to improve statistics. We also find that $\tau$ scales always as $\tau \propto q_{\|}^{-n}$ with $n$ of the order of 1 (figure 8b), a feature common to many arrested systems.

Notably, the shape exponent $\gamma$, shown in panel c) of the same figure, varies as a function of $q_{\|}$ and of the concentration $\Phi$. In the range covered by the present study we are able to document a transition from stretched ($\gamma < 1$) to strongly compressed shape ($\gamma > 1.6$) as the concentration is increased, and $q_{\|}$ is decreased..

The contour line marks the threshold value of $\gamma = 1$, corresponding to simple exponential decay, while the dashed line indicates the threshold concentration for the onset of the percolative cluster. The observed change in shape is a signature of a change in the physical mechanism responsible for the underlying dynamics. While a stretched shape would suggest a distribution of Brownian diffusors, with an heterogeneous distribution of relaxation times, the transition to a compressed shape, accompanied by $n \approx 1$ is a signature of an arrested system. Similar stretched-compressed dynamical crossover have been also reported in few different out of equilibrium materials such as nanoparticles in glass former matrix [35], metallic glasses [36] and, very recently, in Laponite suspensions by Angelini et al. [37]. In the latter, the authors observe $\gamma \approx 0.75$ in an aged Laponite suspension ($Cw = 3.0\%$), while rejuvenation by means of mechanical perturbation seems to induce a more compressed shape ($\gamma = 1.5$ at low $q$, decaying to 1.25 as $q$ increases).

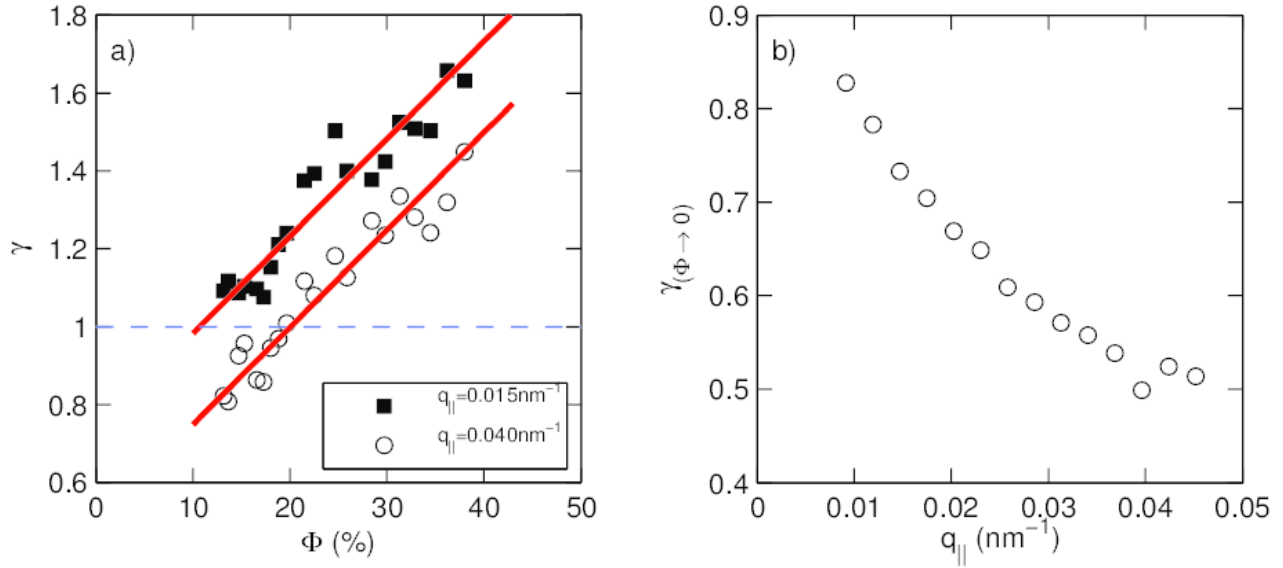

**Figure 6** *(a) the dependence of $\gamma$ on $\Phi$ is linear in the $q_\parallel$ range explored. Continuous lines are linear fit. The dashed line indicates the simple exponential shape ($\gamma = 1$) (b) the extrapolation of $\gamma$ at $\Phi \to 0$ decreases with increasing $q_\parallel$.*

Figure 6a represents a different way of recasting the data of figure 5c, highlighting that, at a given $q_\parallel$, it is possible to control the shape of the correlation functions from stretched to compressed, at will, just by varying the surface concentration, which is achieved simply by moving the barriers of the Langmuir trough. In detail, at every $q_\parallel$ value, we observe a linear relation of $\Phi$ and $\gamma$, with the same slope and varying intercept, whose dependence on $q_\parallel$ is depicted in figure 6b.

The complex dynamics of our system are not trivial to interpret on a microscopic scale. A compressed relaxation is predicted within a detailed model proposed by Bouchaud and Pitard [38] for the dynamics in an elastic solid. Within this model, randomly appearing dipolar stresses generate a field of strains in the network of the elastic gel. It is then assumed that the dynamics of the diffusors is determined by the relaxation of such local strains, leading to the compressed shape of the correlation functions. This detailed model, however, puts strict constraints on the q-dependence of $\gamma$ (predicting two regimes, $\gamma = 1.5$ for small $q$ and $\gamma = 1.25$ for large $q$), which do not agree with our data.

A more phenomenological model was put forward by Duri and Cipelletti some years ago [39]. Those Authors attribute the compressed shape for the relaxation found in DLS and XPCS experiments to rare, intermittent rearrangements. They deduce a variation of the shape from compressed with $\gamma = 1.5$ to simple exponential as a function of the reduced scattering vector, obtained scaling q by a typical length $\delta$ characterizing the displacement. In this case, the lower limit to the shape parameter ($\gamma = 1$) is dictated by the regime in which a single displacement is sufficient to fully decorrelate the signal. The variation of $\gamma$ predicted by this model is manifestly more limited than what we observe. It is possible however that a similar model may hold in our case, if a spatially heterogeneous distribution of relaxation processes generates a distribution of relaxation times, leading to $\gamma < 1$. On the other hand, the upper limit of the range ($\gamma = 1.5$) predicted by this model is connected with the power law decay of the probability distribution function of the displacements $\Delta R$ observed in the gel. This distribution is assumed to be Levy-like [39,40] with its tail towards large $\Delta R$ being proportional to $\Delta R^{-p}$ with $p = \gamma + 1 = 2.5$. Therefore, our results ($\gamma = 1.8$ at low-q and high concentration) would imply a steeper decay of the displacements' probability distribution in the low-q and large $\Phi$ regime, leading to an increased ballistic-like character of the dynamics.

Encouraged by these considerations, we compare our results with a variant of the model proposed in [39]: the values of $\gamma$ measured at each concentration are then plotted in figure 7a, on an adimensional axis provided by the scaled scattering vector $q_\parallel \delta$ so that they overlap, collapsing on a master curve, in analogy with figure 3 of [39]. In this construction, an overall scale factor needs to be determined. This is accomplished by noting that for $q_\parallel \delta = 2$ the model predicts $\gamma = 1$, therefore the absolute values of $\delta$ can be determined. The master curve thus obtained is in agreement with the model curve up to $q_\parallel \delta = 2$, above which value the model would plainly predict $\gamma = 1$, while we observe a stretched shape. As anticipated, this can be reconciled with the model assuming that in our case we observe a spatially heterogeneous distribution of relaxation times. The scaling parameter $\delta$, reported as a function of $\Phi$ in figure 7b, naturally offers us an estimate of the "long time displacement length", i.e. of the dynamics taking place on the time scale of $\tau$. In the same figure, it is compared with the "short time localization length" $r_{loc}$, deduced from the Debye-Waller decay of the contrast, which on the contrary characterizes the length scale of the fast dynamics. While it is notable that a single framework is able to describe the dynamics of this system over a broad range of concentrations, detailed inspection of the figure suggests that two different regimes exist, above and below the percolation threshold: below it, the length scales $r_{loc}$ and $\delta$ of the fast and slow dynamics decouple, while above it they become equal within experimental accuracy, and also comparable to the size of the single nanoparticle.

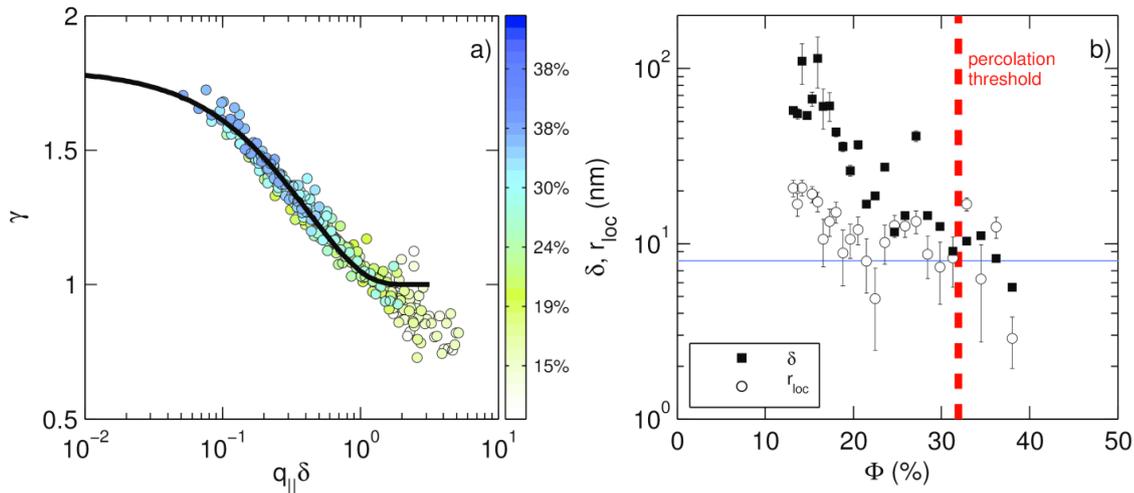

*Figure 7: master curve built from $\gamma$ as a function of scaled scattering vector at different $\Phi$ **a)** $\gamma(q_\parallel)$ curves are shifted along the x-axis by the scaling length $\delta$ so that they overlap. The resulting master curve is in partial agreement with the model (solid line) proposed by Duri et al. [39]. **b)** "long time displacement length" $\delta$ as a function of $\Phi$, compared with the "short time localization length" $r_{loc}$. $\delta$ becomes comparable with the particle size (8nm, horizontal line) and $r_{loc}$ above the percolation threshold (vertical dashed line). Color online.*

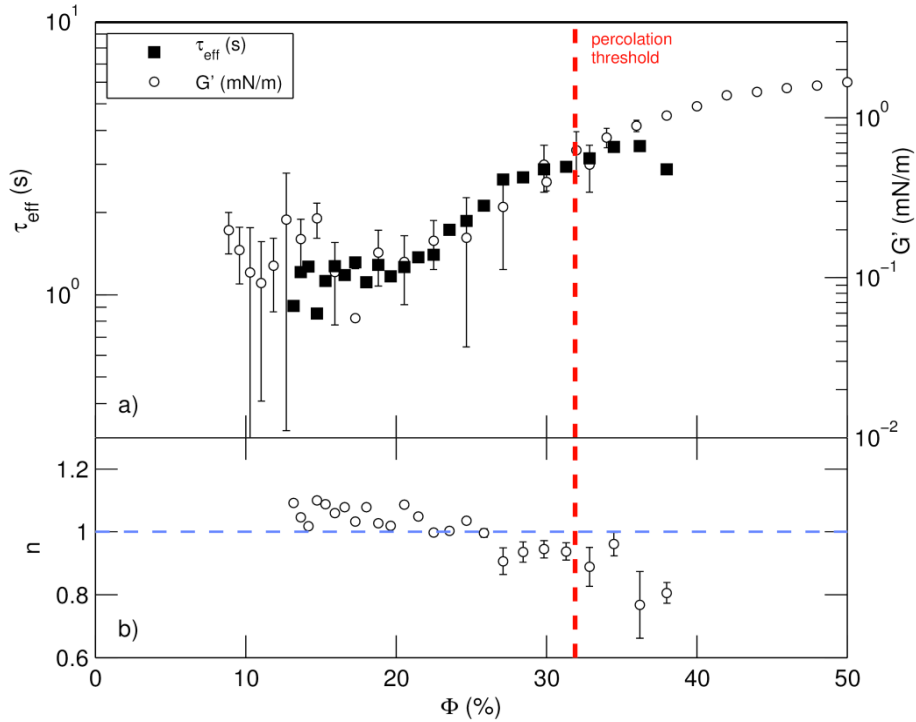

*Figure 8. Approaching the percolation transition, the shear modulus G' and the relaxation time $\tau$ increase following the same law; **a)** comparison of G' and $\tau$ as a function of $\Phi$: they follow the same behavior. Limit values are reached by both quantities above the percolation threshold (vertical red line) **b)** the parameter n in the dependence $\tau \propto q_{\parallel}^{-n}$ decreases as the concentration is increased. The blue dashed line represents $n = 1$. Color online.*

Figure 8 reports the concentration dependence of $\tau$ and of the exponent $n$ detailing the dependence of $\tau$ on $q_{\parallel}$. Defining the 'effective' relaxation time as $\tau_{eff} = \tau \frac{\Gamma(1/\gamma)}{\gamma}$ [41] (where $\Gamma$ is the gamma function) we can compare its behavior with that of the mechanical modulus G. This is shown, for $q_{\parallel} = 0.015 nm^{-1}$, in figure 8a; $\tau_{eff}$ increases as the concentration is raised in the same fashion as the elastic shear modulus $G'$.

In the high concentration regime (e.g. around the percolation threshold) this finding is consistent with the prediction of the already mentioned Bouchaud-Pitard model [38] for the relaxation in an elastic solid. In this case, $\tau$ is predicted to be proportional to the elastic modulus and inversely proportional to $q$ as observed experimentally and reported in figure 8. We remind, however, how the model partly failed in explaining the compressed-to-stretched transition of the correlation functions' shape reported in figure 5c and discussed before.

Notably, we find that the agreement between $\tau_{eff}$ and $G'$ holds on the whole $\Phi$ range, even at those lower concentrations where the correlation function shape is stretched, namely for $\Phi < 18\%$. This may be rationalized considering that, even in this regime, the film dominant mechanical response is elastic. In a similar nano-gel in 3D, Guo et al. [42] verified a relation between G' and localization length which is postulated in a self-consistent mode-coupling theory [43]: this relation does not hold in our case, as $\tau$ is constant for $\Phi < 18\%$ within the error, while the localization length (figure 4b) decreases.

The general picture emerging from the agreement between $\tau_{eff}$ and $G'$ is that of a system in which fluctuation dissipation relations hold, even connecting results spanning very different space scales, from microns (GI-XPCS) to centimeters (rheometry).

Panel 7b reports the dependence of the exponent $n$ on $\Phi$. The coefficient $n \approx 1$ in the scaling $\tau \propto q_{||}^{-n}$ observed by us in the stretched regime ($\Phi < 18\%$) is unexpected and, to the best of our knowledge, it has been reported before only in the case of Laponite. Further theoretical investigations may be needed in order to formulate a physical model for this behavior. On the contrary, the slight reduction of the exponent n towards 0.8 observed in the compressed regime is consistent with what already observed at higher concentration on the same system [18].

### 3.3 Dynamics and relaxation following a perturbation

Following the characterization of the spontaneous dynamics of the 2D network as a function of $\Phi$, we now focus on the effects on the dynamics caused by an external mechanical perturbation due to the oscillatory movement of a magnetic needle floating on the surface, in a similar fashion to the ISR apparatus [24] used to measure the mechanical modulus of the system, as better described in the experimental section.

To check for stationarity, and to better characterize each starting point, the dynamics of the system has been measured before applying any perturbation; then the perturbation was applied for 10 minutes, with the magnetic needle put into angular oscillation on the surface, at frequency of 10Hz and amplitude $\alpha \approx 5°$. The needle was $4\ cm$ long: given the total area occupied by the film, this perturbation induced a relative area variation $\frac{dA}{A} \approx 0.3\%$. The perturbation ended by bringing the needle back to its initial position. In this way, we can exclude that the dynamics observed after the perturbation reflect some obvious stress relaxations on the macro-scale.

During the needle's movement, no GI-XPCS measurement was possible, due to the obvious and drastic decorrelation induced by the mechanical perturbation.

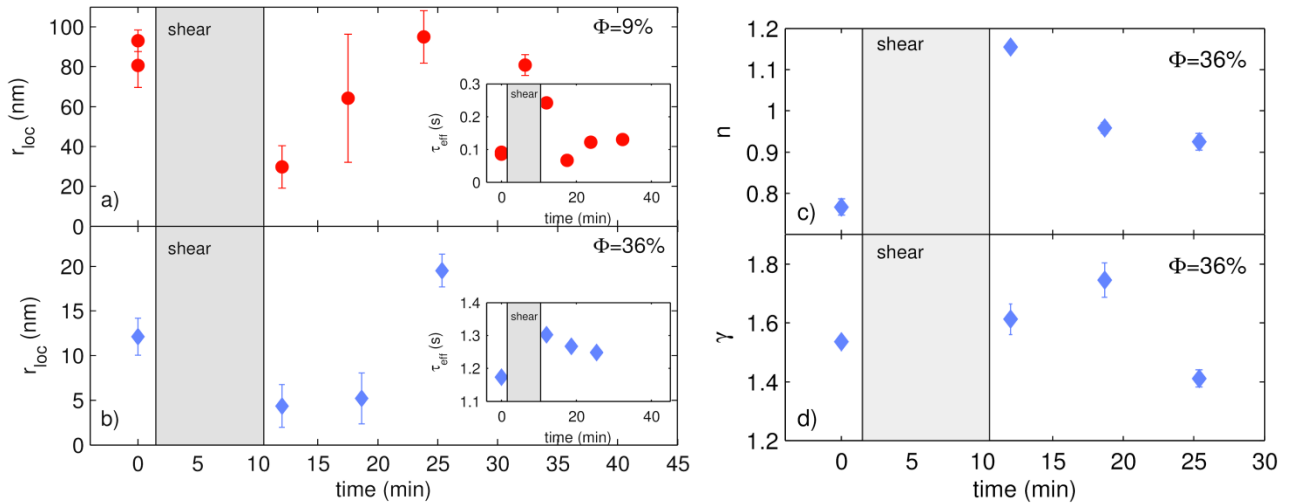

**Figure 9: A mechanical perturbation affects the fluctuation dynamics in different ways for films above and below the percolation threshold.** *The localization length, measured below (9%, **a**) and above (36%, **b**) the percolation threshold. While a major effect of compaction is induced at 9%, very little effect is produced at 36%; alongside this, the recovery of the unperturbed state is much quicker at low concentration. Insets: effective relaxation times increase after perturbation, measured at $q_{||} = 0.02 nm^{-1}$. **c, d)** At $\Phi = 36\%$, the mechanical perturbation causes and increase in the slope of the $\tau \propto q_{||}^{-n}$ dependence and in the shape exponent $\gamma$. Color online.*

After switching off the perturbation, the dynamics of the network was measured following the usual procedure, mapping the evolution of the dynamical parameters extracted from the correlation functions as a function of the experimental time, on a temporal scale of tens of minutes.

The experiment was performed at two different concentrations- namely at $\Phi = 9\%$ and $\Phi = 36\%$ - chosen to be below and above the percolation threshold; the results are reported in figure 9a and 9b. The first, striking result is that in both cases the localization length $r_{loc}$ is found to be sensibly reduced by the perturbation, and its unperturbed value is recovered only on a time scale as slow as 10-20 minutes, though the recovery seems to be faster at low concentration. Alongside this, $\tau$ is increased; at concentration $\Phi = 9\%$ it relaxes back to values close to those of the system's unperturbed state in less than 10 minutes, while at concentration $\Phi = 36\%$, the effect of the perturbation on the dynamics is persistent on a much longer time scale, indicating that a relevant permanent modification of the 2D network might have been induced. Consistently, for $\Phi = 36\%$, the parameter n shows an increase from $n \simeq 0.8$ (unperturbed value) up to $n \simeq 1.2$ immediately after the perturbation, to slowly decrease in the subsequent times towards the unperturbed value (figure 9c).

More relevant changes are found in the analysis of the shape parameter (figure 9d). The perturbation induces a more compressed form, reaching $\gamma \simeq 1.8$ $at$ $q_{||} = 0.02 \text{nm}^{-1}$, consistent with the interpretation of the data provided in [43] for Laponite. Notably, this stress-induced effect was not observed instantaneously at the switch off of the perturbation, rather it peaks roughly after 10 minutes after this. For longer times, the system then reaches a new state characterized by $\gamma \simeq 1.4$, a value slightly lower than the unperturbed state, that may be interpreted as a new stationary state in which stress relaxation is comparably less important than before the perturbation was applied.

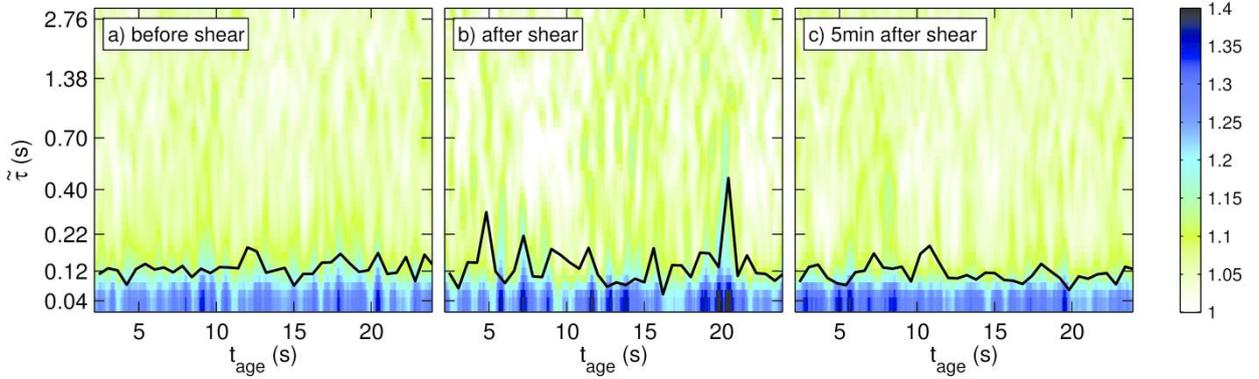

*Figure 10: insight on the effects of the mechanical stress on the dynamics, displaying the age dependence of the correlation functions, $g^{(2)}(t_{age}; \tilde{\tau})$ at $\phi = 9\%$ and $q_{||} = 0.02 nm^{-1}$. Black lines represents $\tau$ as a function of the ageing time. The mechanical stress induces heterogeneities in the characteristic time of the dynamics, that disappear in a few minutes. Color online.*

In order to investigate the temporal evolution of the dynamics and to shed more light on transient effects, we calculate the two-times correlation functions, defined as

$$C(t_1, t_2) = \frac{\langle I(t_1)*I(t_2)\rangle}{\langle I(t_1)\rangle\langle I(t_2)\rangle} \text{ eq.(7)}$$

where the average is performed over the pixels characterized by the same $q_{||}$ and $I(t_1)$ and $I(t_2)$ are the scattered intensities measured at two different experimental times $t_1$ and $t_2$. It is customary to represent it as functions of the ageing time $t_{age} = |t_2 + t_1|/2$ and of the lag time

$\tilde{\tau} = |t_2 - t_1|$; graphically, one can imagine to extract $g^{(2)}(t_{age}; \tilde{\tau})$ correlation functions by selecting rectangular slices of $C(t_1, t_2)$ taken perpendicular to its main diagonal [14].

The calculation is performed on data measured before and after the application of the stress, $q = 0.02 \text{nm}^{-1}$ and for concentration $\Phi = 9\%$. Subsequently, the correlation functions $g^{(2)}(\tilde{\tau})|_{t_{age}}$ selected at different values of $t_{age}$ have been fitted with simple exponential decays, in order to obtain the evolution of the relaxation time $\tau$ with increasing $t_{age}$.

Figure 10 shows $g^{(2)}(t_{age}; \tilde{\tau})$ measured before and after the mechanical perturbation: relaxation times $\tau(t_{age})$ resulting from the fitting procedure is plotted as black lines.

The application of the mechanical stress causes the onset of a dynamical heterogeneous state, with successive random appearance of slow dynamics, even after that the mechanical perturbation has been switched off. This causes, on average, the slower relaxation time extracted from the correlation functions $g^{(2)}(\tau)$, reported in the inset of figure 9a. The dynamical heterogeneous state disappears after few minutes, with the complete recovery of the unperturbed state.

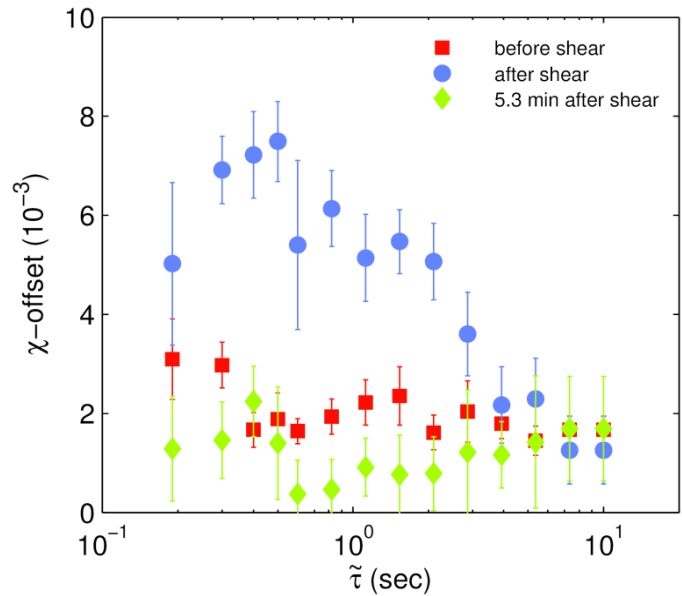

*Figure 11: effects of the mechanical stress on the variance $\chi$ of the two-times correlation functions, at $\phi = 9\%$. The mechanical stress induces an increase of the variance of the two times correlation functions, with the appearance of a peak. This peak disappears after a waiting time of a few minutes (flat variance). Color online.*

This is confirmed by the analysis of the variance $\chi(\tilde{\tau})$ of the two-times correlation function, defined as:

$$\chi(\tilde{\tau}) = \langle C^2(t_{age}; \tilde{\tau}) \rangle - \langle C(t_{age}; \tilde{\tau}) \rangle^2 \quad \text{eq.(8)}$$

which is also related to the dynamical susceptibility.

As shown in figure 11, in the unperturbed state the system is characterized by a featureless variance; after the mechanical perturbation, the variance shows a peak centered at $\tilde{\tau} \approx 0.8 \, s$, which disappears at longer times, when the effects of the perturbation vanish. The presence of a peak in the dynamical susceptibility is commonly associated, in the literature, with the onset of dynamical heterogeneity; in our case, it may indicate that the mechanical perturbation affects the 2D network in a non-uniform way, activating random relaxing stresses that propagate as density fluctuations and disappear in the time range of a few minutes. This phenomenology is consistent with analogous results obtained by means of DLS measurements on colloidal gel [39].

## 4. CONCLUSIONS

The evolution of the dynamics of a bidimensional gel formed by gold nanoparticles at the air/water interface has been characterized by means of GI-XPCS, supported by imaging and by rheology measurements. Mechanical measurements have shown that the structure behaves as a bidimensional elastic solid at all concentrations. The increase of the elastic shear modulus as a function of surface concentration closely mimics that of the GI-XPCS relaxation time. The general picture emerging is that of a system in which fluctuation dissipation relations hold, even connecting results spanning very different space scales, from microns (GI-XPCS) to centimeters (rheometry). In particular, this has been predicted within a model proposed by Bouchaud and Pitard [38], which however fails to account for the observed shapes of the correlation functions: we document in our system the first example, to the best of our knowledge, of an actively controlled transition from stretched to compressed shape. The observed behavior seems better reconcilable with a phenomenological model proposed by Duri and Cipelletti [39] which hypothesizes that the correlation of the scattered intensity is driven by rare intermittent rearrangements. Exploiting the unique opportunity offered by the Langmuir 2D geometry to continuously vary the concentration on the very same gel system, we could push further this analysis and we have built a novel master-curve for the shape parameter as a function of the reduced scattering vector. According to the model [39], the scaling factor of this master curve corresponds to the "long time displacement length" $\delta$, characterizing the dynamics happening on the time scale of a few seconds. This is distinct from the "short time localization length $r_{loc}$, determined by the usual pseudo Debye-Waller analysis of the initial contrast, which characterizes dynamics faster than 10 milliseconds. In the low concentration regime we find the $\delta \simeq 100\ nm$, much larger than the value of $r_{loc}$; however as the concentration $\Phi$ increases towards the percolation threshold, the two lengths converge to a common limit comparable to the single particle size.

We also address dynamical heterogeneities, extending our previous observations [18] by focusing here on the low concentrations and on the deep out-of-equilibrium regimes induced by an external mechanical perturbation. This perturbation induces an increase of the relaxation time, accompanied by an increase of the shape exponent $\gamma$. This fact indicates a more compressed, ballistic-like character of the dynamics as a consequence of the perturbation applied, and suggests an analogy with the results obtained by XPCS on aged and rejuvenated Laponite [43].
At the same time, perturbation enhances the temporal heterogeneities, leading to the appearance of a peak in the variance. After the end of the perturbation, the relaxation towards the initial stationary state, happens on the time scale of a few minutes at low concentration, while at high concentration takes much longer.

### Acknowledgements and Author contributions

The ESRF is gratefully acknowledged for provision of beam time; O. Konovalov (ESRF) for access to the Langmuir laboratory. This work is partly held under the umbrella of the COST CM1101 and MP1106 Actions.

D.O. and L.C. conceived the project. D.O., B.R., Y.C., G.B., T.R. and L.C. performed the XPCS experiments; A.P. and G.R. prepared the samples; D.O., T.R. and L.C. analyzed the data. L.C. and D.O. wrote the manuscript. All the Authors discussed the data and commented on the manuscript.